\documentclass[journal]{igtesymp}
\ifCLASSINFOpdf
\else
\fi
\usepackage[tight,footnotesize]{subfigure}

\usepackage{amsmath}
\usepackage{amssymb}
\usepackage{graphicx}
\usepackage[utf8]{inputenc}
\usepackage{umlaute}

\begin{document}
%
\title{Magnetic Field Simulations Using Explicit Time Integration With Higher Order Schemes}
%
%
%
\author{Bernhard~Kähne\IEEEauthorrefmark{1},  Markus Clemens\IEEEauthorrefmark{1}
and Sebastian Schöps\IEEEauthorrefmark{2} 
 \\ \vspace{0.3cm} \normalsize{
\IEEEauthorblockA{\IEEEauthorrefmark{1}University of Wuppertal, Chair for Electromagnetic Theory, Rainer-Gruenter-Str. 21, 42119 Wuppertal, Germany\\
 \IEEEauthorrefmark{2} Graduate School of Computational Engineering, Technical University of Darmstadt, 64293 Darmstadt, Germany\\
 E-mail: kaehne@uni-wuppertal.de}}
}

%
%


\IEEEaftertitletext{\vspace{-1cm}\noindent\begin{abstract}
A transient magneto-quasistatic vector potential formulation involving nonlinear material
is spatially discretized using the finite element method of first and second polynomial order.
By applying a generalized Schur complement the resulting system of differential algebraic
equations is reformulated into a system of ordinary differential equations (ODE).
The ODE system is integrated in time using the explicit Euler scheme, which is conditionally stable
by a maximum time step size. To overcome this limit, an explicit multistage Runge-Kutta-Chebyshev time integration method
of higher order is employed to enlarge the maximum stable time step size. Both time integration 
methods are compared regarding the overall computational effort.


\end{abstract}
\noindent\vspace{\baselineskip}}

\maketitle
\thispagestyle{empty}\pagestyle{empty}

%
\IEEEpeerreviewmaketitle

\section{Introduction}
The simulation of magneto-quasistatic (MQS) field problems is crucial in the design process
of electric machines, transformers and any other devices in the low frequency regime. Usually,
these applications have material coefficients, which differ by many magnitudes through the
computational domain like e.g. the change of the conductivity through the rotor,
air-gap and stator of an electric engine. Furthermore, often some conductive parts of
the problem are ferromagnetic with a nonlinear B-H-characteristic.

Spatial discretization of MQS field problems based on a magnetic vector potential formulation
using the finite element method (FEM) yields nonlinear infinitely
stiff differential-algebraic equation (DAE) systems of index 1. This DAE is commonly integrated in time
employing implicit integration techniques \cite{Hairer-Wanner}. A nonlinear system of equations 
has to be solved by e.g. the Newton-Raphson method, which requires multiple iterations and
repeated calculations of the Jacobian matrix in order to carry out one implicit time step. This is
computationally expensive, but, due to implicit integration, unconditionally stable with no
restriction of the time step size.

In this paper, the DAE system is translated into a system of ordinary differential equations (ODE) using
the generalized Schur complement, as was originally presented in \cite{Cle-Scho-DeG}. This ODE system is
integrated in time by employing the explicit Euler scheme, which is conditionally stable by a
maximum time increment according to a Courant-Friedrichs-Levi (CFL) -like criterion. Unlike
in the implicit integration scheme, here the use of a linearization scheme is avoided. Hence, one
explicit time step is computationally less expensive. Based on previous numerical investigations
presented in \cite{DutineClem2017} and \cite{DutineClemSchoep2017}, this paper presents
numerical results for higher order
spatial discretizations. Furthermore, a explicit multistage Runge-Kutta-Chebyshev (RKC) method
of higher order is employed in order to incease the maximum stable time step size as presented in
\cite{Hairer-Wanner} and \cite{Sommeijer}. The RKC method provides to control the maximum stable
time step size which is quadratic in the number of stages.

\section{Mathematical Formulation}
The magnetic flux density $\vec B$ is described by the magnetic vector potential $\vec A$ such that
\begin{equation}
B = \nabla\times\vec A.
\end{equation}
Using the non-gauged magnetic vector potential in $\vec A^*$-formulation, the governing MQS boundary value
problem for the computational domain $\Omega$, see Fig. \ref{fig:domains}, reads
\begin{equation}
\label{eq:mqs_formulation}
\begin{split}
\nabla\times(\nu \nabla\times\vec A(\vec r,t)) + \kappa\frac{\partial\vec A(\vec r,t)}{\partial t}
&= \vec J(\vec r,t) \quad \forall \vec r \in \Omega,\\ 
\vec n \times \vec A(\vec r,t) &= 0 \quad \forall\vec r \in \partial\Omega
\end{split}
\end{equation}
where $\nu$ is the nonlinear
reluctivity, typically dependent on $\vec B$, $\kappa$ is
the electrical conductivity and $\vec J$ is current density.
The corresponding variational formulation of (\ref{eq:mqs_formulation}) reads:
find
$\vec A\in \mathcal H_0(\mathbf{curl}, \Omega)$ such that
\begin{equation}
\label{eq:varformMQS}
\begin{split}
\int_{\Omega} \kappa\frac{\partial\vec A}{\partial t}\cdot\vec w\;\mathrm d\Omega
+\int_{\Omega}\nu (\nabla\times\vec A)\cdot (\nabla\times\vec w)\;\mathrm d\Omega\\
= \int_{\Omega}\vec J\cdot\vec w\;\mathrm d\Omega
\end{split}
\end{equation}
holds true for all curl-conforming edge element test functions
$\vec w\in \mathcal H(\mathbf{curl}, \Omega)$.

\subsection{Spatial discretization}
\begin{figure}[]
\centerline{\includegraphics[width=0.75\columnwidth]{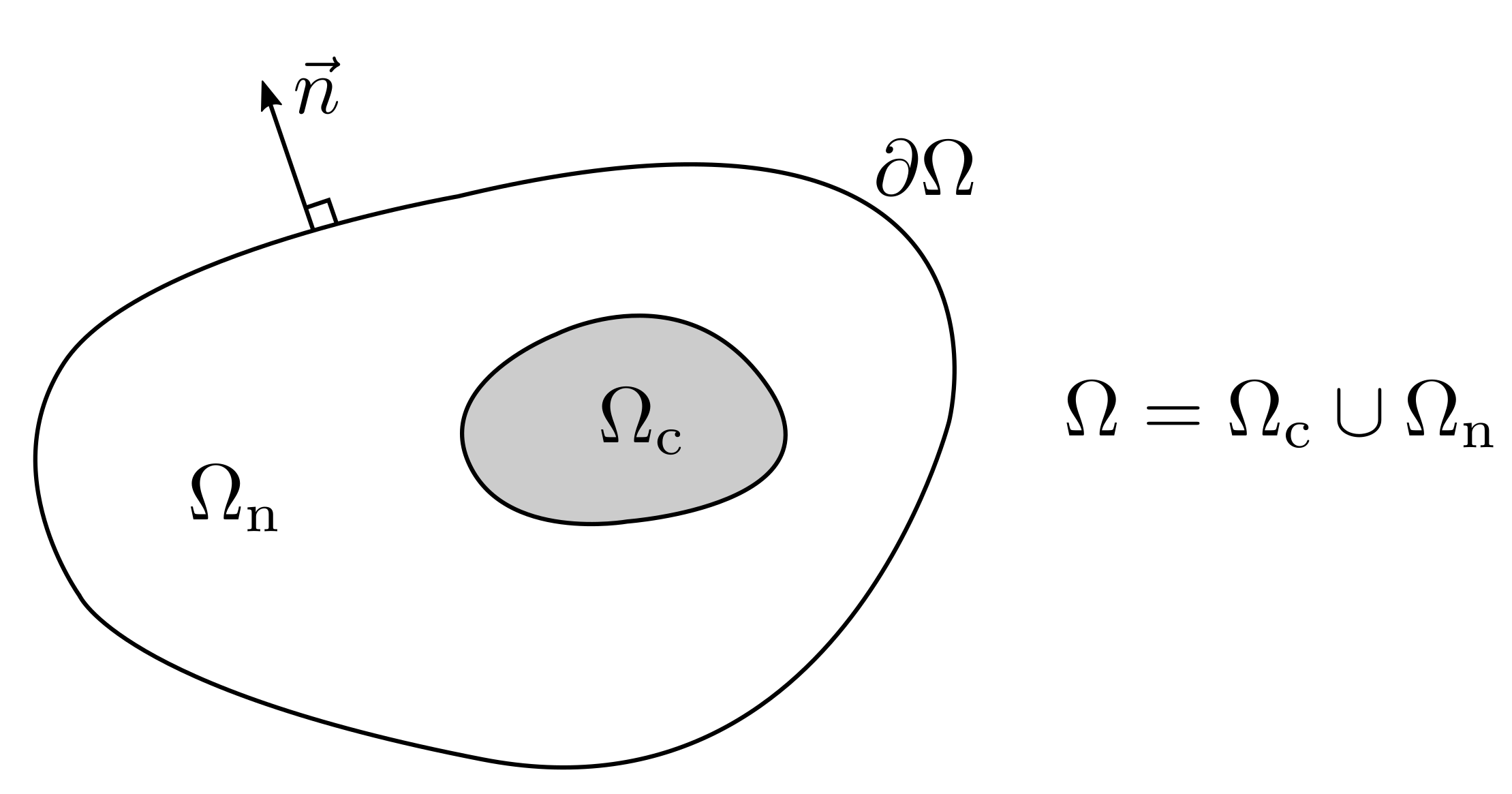}}
\caption{Computational domain $\Omega$ splitted into a conductive $\Omega_\mathrm{c}$
and a non-conductive $\Omega_\mathrm{n}$ subdomain with
boundary $\partial\Omega$ and outer normal vector $\vec n$.}
\label{fig:domains}
\end{figure}
The vector potential
$\vec A$
is approximated by a finite set of edge elements:
\begin{equation}
\vec A\approx \vec A_h=\sum^{N}_{i=1} a_i\vec w_i,\quad
\end{equation}
where $\vec w_i \in \mathcal H_0(\mathbf{curl}, \Omega)$ and $N$ denotes the number
degrees of freedom (DOF). According to the Galerkin scheme the test functions
are taken from the same basis as $\vec A_h$.
Coefficients $a_i$
assigned to the conductive area $\Omega_\mathrm{c}$ are stored in vector $\boldsymbol a_{\mathrm c}$,
likewise vector $\boldsymbol a_{\mathrm n}$ holds the coefficients in the
nonconducting  area $\Omega_\mathrm{n}$, see Fig. \ref{fig:domains}. It follows the DAE system
\begin{equation}
\label{eq:mqs-dae}
\left[\begin{matrix}
 \boldsymbol M_{\mathrm c} & 0\\ 0 & 0
\end{matrix}\right]
\frac{\mathrm d}{\mathrm d t}
\left[\begin{matrix}
 \boldsymbol a_{\mathrm c}\\  \boldsymbol a_{\mathrm n}
\end{matrix}\right]
+
\left[\begin{matrix}
\boldsymbol K_{\mathrm c}(\boldsymbol a_{\mathrm c}) & \boldsymbol K_{\mathrm{cn}}\\
\boldsymbol K^{\mathsf T}_{\mathrm{cn}} & \boldsymbol K_{\mathrm n}
\end{matrix}\right]
\left[\begin{matrix}
\boldsymbol a_{\mathrm c}\\ \boldsymbol a_{\mathrm n}
\end{matrix}\right]
=
\left[\begin{matrix}
0\\ \boldsymbol j
\end{matrix}\right]
\end{equation}
where $\boldsymbol M_{\mathrm c} $ is the positive-definite and symmetric conductivity matrix, $\boldsymbol K_{\mathrm c}$ is the time dependent
curl-curl matrix in the conductive region and $\boldsymbol K_{\mathrm n}$ is the constant curl-curl
matrix in the non-conductive area. $\boldsymbol K_{\mathrm{cn}}$ is the coupling matrix, which
couples the DOF in the conductive area with neighboring DOF in
the non-conductive area. The vector $\boldsymbol j$ defined by the right-hand side (RHS)
of the variational formulation (\ref{eq:varformMQS}) is only present to the non-conductive
region  $\Omega_\mathrm{n}$.
Due to the null-space of the
discrete curl-curl operator $\boldsymbol K_{\mathrm n}$, a vanishing weak divergence of $\boldsymbol j$ in $\Omega_\mathrm{n}$
is crucial for the convergence of iterative solvers having a weak gauging property, e.g. the preconditioned 
conjugate gradient (PCG) method \cite{ClemWeil1999}.

Applying the generalized Schur complement
\begin{equation}
\label{eq:schur}
\boldsymbol K_{\mathrm S}(\boldsymbol a_{\mathrm c}) = \boldsymbol K_{\mathrm c}(\boldsymbol a_{\mathrm c}) - \boldsymbol K_{\mathrm{cn}}
\boldsymbol K^+_{\mathrm n}\boldsymbol K^{\mathsf T}_{\mathrm{cn}}
\end{equation}
where $\boldsymbol K^+_{\mathrm n}$ is the Moore-Penrose pseudo-inverse matrix of the singular curl-curl matrix $\boldsymbol K_{\mathrm n}$, equation
(\ref{eq:mqs-dae}) decomposes into a finitely stiff ODE system and an algebraic
equation given by
\begin{eqnarray}
\label{eq:ode}
\boldsymbol M_{\mathrm c}\frac{\mathrm d}{\mathrm d t}\boldsymbol a_{\mathrm c} + \boldsymbol K_{\mathrm S}(\boldsymbol a_{\mathrm c})
\boldsymbol a_{\mathrm c} &=& - \boldsymbol K_{\mathrm{cn}} \boldsymbol K^+_{\mathrm n}\boldsymbol j\\
\label{eq:algEq}
\boldsymbol a_{\mathrm n} &=& \boldsymbol K^+_{\mathrm n}\boldsymbol j -  \boldsymbol K^+_{\mathrm n}
\boldsymbol K^{\mathsf T}_{\mathrm{cn}} \boldsymbol a_{\mathrm c}.
\end{eqnarray}
Finally, the system of ODEs (\ref{eq:ode}) has the canonical form
\begin{equation}
\label{eq:canonicalODE}
\begin{split}
\frac{\mathrm d}{\mathrm d t}\boldsymbol a_{\mathrm c}(t)
=\underbrace{-\boldsymbol M^{-1}_{\mathrm c}[ \boldsymbol K_{\mathrm{cn}}
\boldsymbol K^+_{\mathrm n}\boldsymbol j +\boldsymbol K_{\mathrm S}(\boldsymbol a_{\mathrm c})
\boldsymbol a_{\mathrm c}]}_{=:\boldsymbol f(t, \boldsymbol a_{\mathrm c}(t))}.
\end{split}
\end{equation}

\subsection{The explicit Euler time integration scheme}
Replacing the time derivative in ODE system (\ref{eq:canonicalODE}) by the forward
discrete difference yields to the explicit Euler scheme. The solution for
the $(i+1)$-th time step reads
\begin{equation}
\boldsymbol a^{i+1}_{\mathrm c} = \boldsymbol a^i_{\mathrm c}
+ \tau  \boldsymbol f(t^i, \boldsymbol a^i_{\mathrm c}),
\end{equation}
where $\tau = t^{i+1}-t^i$ is the time step size. The maximum stable
time step size is determined by
\begin{equation}
\label{eq:exlpEulerStab}
\tau \leq \frac{2}{\lambda_{\mathrm{max}}(\boldsymbol M^{-1}_{\mathrm c}
\boldsymbol K_{\mathrm S}(\boldsymbol a_{\mathrm c}))},
\end{equation}
where $\lambda_{\mathrm{max}}$ is the maximum eigenvalue.

\begin{figure}[]
\centerline{\includegraphics[width=\columnwidth]{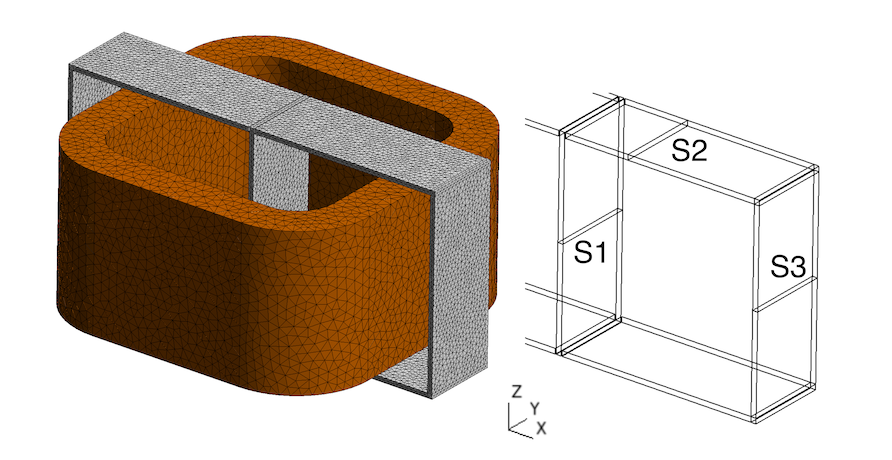}}
\caption{Left: structure and tetrahedral mesh of the TEAM 10 benchmark problem with
about $280\,000$ elements. Symmetrically placed steel plates (gray) around a coil (brown).
Right: cross section S1-S3 of the steel plates with the dimension
$3.2\,\mathrm{mm}\times 50\,\mathrm{mm}$.}
\label{fig:structure}
\end{figure}

\subsection{The Runge-Kutta-Chebyshev  time integration method}
Alternatively, the system of ODEs (\ref{eq:canonicalODE}) can be integrated explicitly in time
by using the RKC method with a number of stages $s$. In order to
compute the $(i+1)$-th time step the stages have the form
\begin{equation}
\label{eq:RKC}
\begin{split}
\boldsymbol y_0 &=\boldsymbol a^i_{\mathrm c}\\
\boldsymbol y_1 &= \boldsymbol y_0 + \tilde\mu_1\tau\boldsymbol f^i_0\\
\boldsymbol y_j &= (1-\mu_j-\nu_j)\boldsymbol y_0 + \mu_j\boldsymbol y_{j-1}
+ \nu_j\boldsymbol y_{j-2}\\
&\quad+ \tilde\mu_j \tau\boldsymbol f^i_{j-1} + \tilde\gamma_j\tau\boldsymbol f^i_{0},
\quad j=2,...,s\\
\boldsymbol a^{i+1}_{\mathrm c} &= \boldsymbol y_s,
\end{split}
\end{equation}
with $\boldsymbol f^i_j = \boldsymbol f(t^i+c_j\tau, \boldsymbol y_j)$.
All coefficients appearing in the RKC scheme (\ref{eq:RKC}) are given analytically
in \cite{Sommeijer} for an arbitrary number of stages $s\geq 2$. Let $T_j$ be the
Chebyshev polynomial of the first kind and degree $j$. For $2\leq j\leq s$ the coefficients are
\begin{equation}
\begin{split}
\tilde\mu_1 = b_1 w_1,\quad \mu_j=\frac{2b_j w_0}{b_{j-1}},\quad \nu_j=\frac{-b_j}{b_{j-2}},\\
\tilde\mu_j=\frac{2b_j w_1}{b_{j-1}},\quad \tilde\gamma_j=-(1-b_{j-1}T_{j-1}(w_0))\tilde\mu_j,
\end{split}
\end{equation}
with
\begin{equation}
\begin{split}
\label{eq:coeff}
\varepsilon = \frac{2}{13},\quad
\quad w_0=1+\frac{\varepsilon}{s^2},\quad w_1=\frac{T^{\prime}_s(w_0)}{T^{\prime\prime}_s(w_0)},\\
\quad b_j=\frac{T^{\prime\prime}_j(w_0)}{(T^{\prime}_j(w_0))^2}, \quad b_0=b_2,\quad b_1=b_2
\end{split}
\end{equation}
and
\begin{equation}
\begin{split}
c_j = \frac{T^{\prime}_s(w_0)}{T^{\prime\prime}_s(w_0)}\frac{T^{\prime\prime}_j(w_0)}{T^{\prime}_j(w_0)}\quad (2\leq j\leq s-1)\\
c_1 = \frac{c_2}{T^{\prime}_2(w_0)},\quad c_s=1.
\end{split}
\end{equation}
The condition for absolute stability is given by
\begin{equation}
\label{eq:RKCStab}
\tau\leq 0.653 \cdot\frac{s^2}{\lambda_{\mathrm{max}}(\boldsymbol M^{-1}_{\mathrm c}
\boldsymbol K_{\mathrm S}(\boldsymbol a_{\mathrm c}))}.
\end{equation}
Unlike the stability criterion (\ref{eq:exlpEulerStab}) of the explicit Euler scheme,
the maximum stable time increment increases quadratically in the number of
stages $s$. 

\section{Numerical Case Study}

\begin{figure}[]
\centerline{\includegraphics[width=\columnwidth]{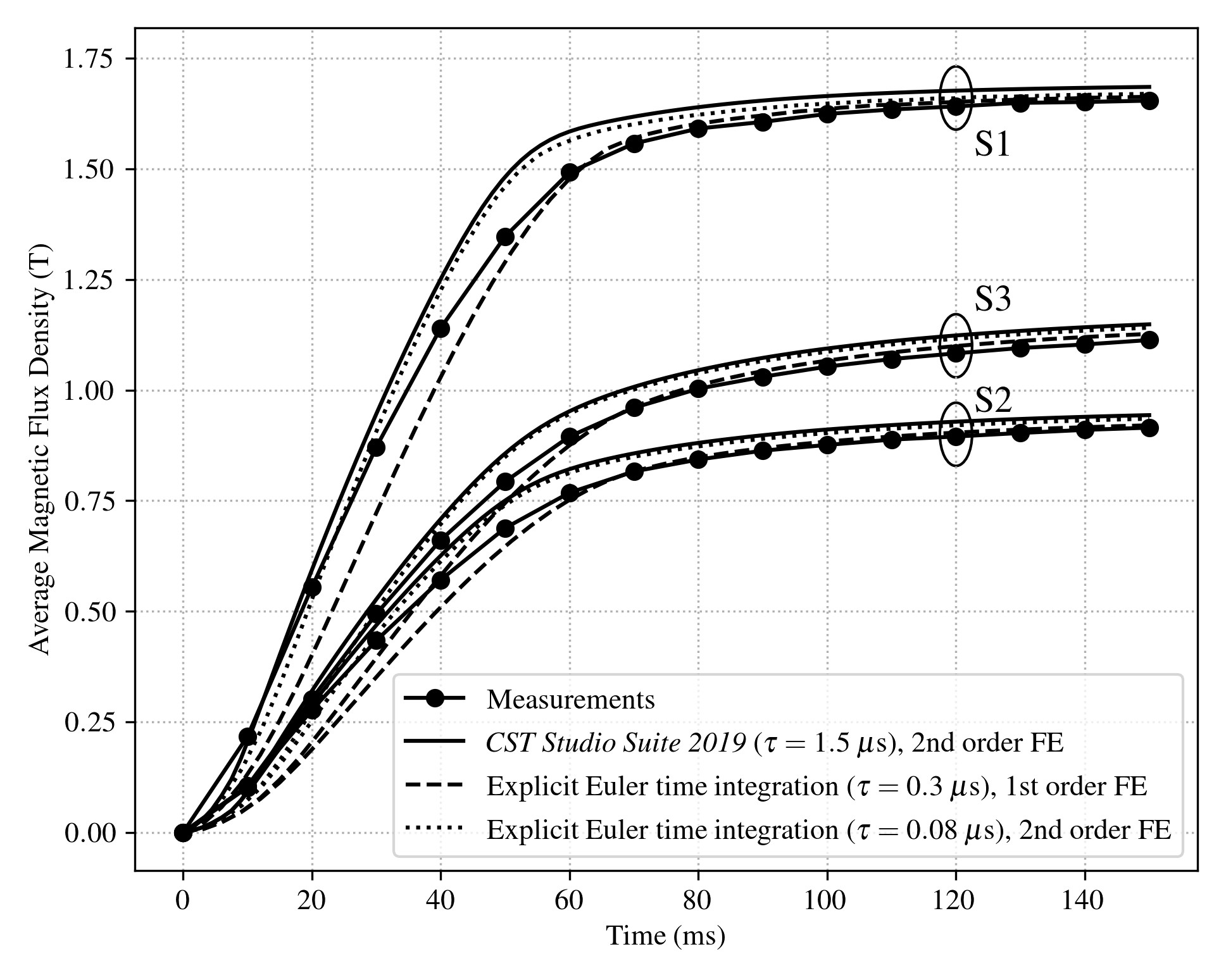}}
\caption{Results for the average magnetic flux density evaluated at the cross sections S1-S3,
see Fig. \ref{fig:structure}, compared with measurements from \cite{NakataFuji1990}. Additional
validation with unconditionally stable implicit time integration using \emph{CST Studio Suite 2019}
software \cite{CST} on a different tetrahedral mesh with about $2\,000\,000$ elements.}
\label{fig:plot_mag_flux}
\end{figure}

The case study in order to validate the formulations above is the TEAM 10 benchmark
problem presented in \cite{NakataFuji1990}. It contains symmetrically placed steel
plates around a coil, see Fig. \ref{fig:structure}. At the junction from the center
plate to the left and right 
canal exists air gaps of $0.5\,\mathrm{mm}$. Furthermore, the conductive steel plates
have a nonlinear B-H-characteristic. The coil has 162 turns and is excited by a
transient current
\begin{equation}
\label{eq:excitation}
i_\mathrm{S}(t) = 5.64\,\mathrm A \cdot \left( 1-\mathrm e^{-\frac{t}{0.5\,\mathrm s}} \right).
\end{equation}
Since the coil is modeled as a stranded conductor, the source current density appearing
on the RHS of (\ref{eq:mqs_formulation}) is described as $\vec J(\vec r, t)=\vec X_{\mathrm{S}}(\vec r)i_\mathrm{S}(t)$,
where $\vec X_{\mathrm{S}}$ is the spatial current distribution. Is is computed using an electric
vector potential formulation presented in \cite{Ren1996}.


In order to evaluate the matrix-vector product with the pseudo-inverse of the matrix $\boldsymbol K_{\mathrm n}$
in the RHS of ODE (\ref{eq:canonicalODE}), a parallel PCG method with Jacobi preconditioner is
applied. Since the matrix  $\boldsymbol M_{\mathrm c}$ has full rank, the corresponding system of equations can
be solved using a parallel direct solver. The evaluation of the vector potential
$\boldsymbol a_{\mathrm c}$ in the conductive region is independent of the vector potential
$\boldsymbol a_{\mathrm n}$ in the non-conductive
region. Thus, the computation of $\boldsymbol a_{\mathrm n}$ according to equation 
(\ref{eq:algEq}) can be considered as a post-processing step. Furthermore, the matrix vector product
$\boldsymbol K^+_{\mathrm n}\boldsymbol j$ is essentially a linear magneto-static problem in the non-conductive
area. Thus, it is computed only once with $i_{\mathrm S}=1\,\mathrm A$ and scaled with $i_{\mathrm S}(t)$
by evaluating the canonical RHS  $\boldsymbol f$ at time $t$.

Since $\boldsymbol f$ is evaluated in each time step at least once, the repetitive evaluation of
the pseudo-inverse $\boldsymbol K^+_{\mathrm n}$ forms a multiple right-hand side problem, \cite{DutineClem2017}.
Therefore, the proper orthogonal decomposition (POD) is used in order to generate improved initial vectors 
for the PCG method.

\subsection{Update strategy of the time dependent curl-curl matrix}
\label{sec:update}
The nonlinear reluctivity $\nu(|\nabla\times\vec A|)$ involves an update of the curl-curl matrix
$\boldsymbol K_{\mathrm c}(\boldsymbol a_{\mathrm c})$. Due to explicit time integration the 
curl-curl matrix is evaluated for past instances of the vector potential $\boldsymbol a_{\mathrm c}$.
Hence, a linearization scheme is avoided. However,
the Runge-Kutta stages (\ref{eq:RKC}) still require multiple evaluations of $\boldsymbol K_{\mathrm c}$
per time step which are costly. This approach may therefore become inefficient since the maximal time step size , i.e.
(\ref{eq:exlpEulerStab}) and (\ref{eq:RKCStab}), does not reflect the dynamics of the solution.
We suggest to bypass evaluations of the nonlinearity material curve.

As proposed in \cite{DutineClemSchoep2017} the matrix $\boldsymbol K_{\mathrm c}(\boldsymbol a_{\mathrm c})$
is kept constant for several time steps as long as
\begin{equation}
\label{eq:update}
 \frac{||\boldsymbol a^*_{\mathrm c}-\boldsymbol a^i_{\mathrm c} ||}{||\boldsymbol a^*_{\mathrm c}||}
 \leq tol
\end{equation}
holds true for a given tolerance $tol$, where $\boldsymbol a^i_{\mathrm c}$ is the latest solution
and $\boldsymbol a^*_{\mathrm c}$ denotes the solution at the latest matrix update.
In this case study the matrix $\boldsymbol K_{\mathrm c}(\boldsymbol a_{\mathrm c})$ is
updated if $tol$ exceeds $0.5$\%.

\begin{figure}[]
\centerline{\includegraphics[width=\columnwidth]{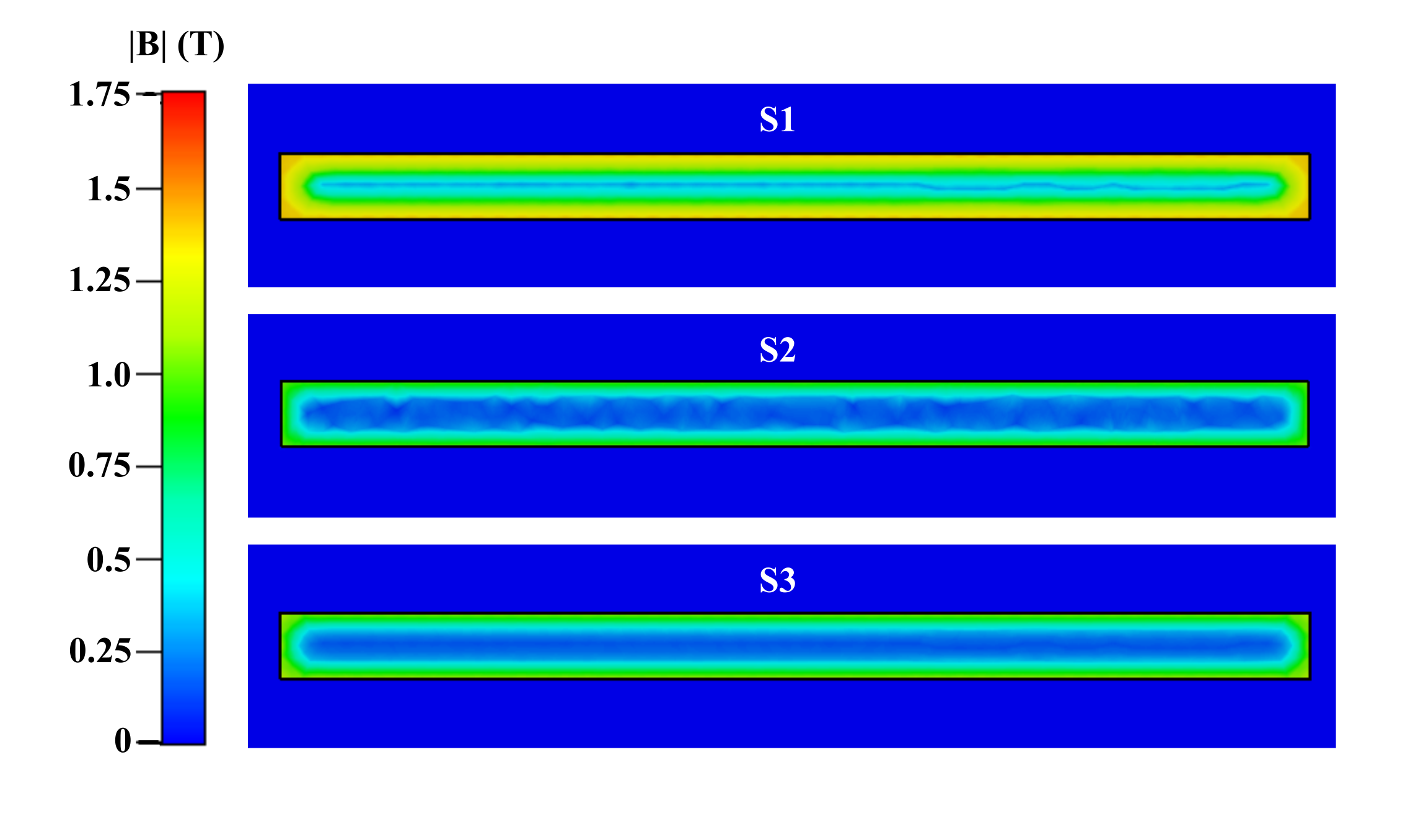}}
\caption{Illustration of the skin effect inside the conductive parts.
The magnetic flux density $\vec B$ on cross sections S1-S3, see Fig. \ref{fig:structure}. 
Second order FEM solution at time $t=30\;\mathrm{ms}$ obtained from \emph{CST Studio Suite 2019}.}
\label{fig:plt_skin_effect}
\end{figure}

\subsection{Numerical results}
The FEM simulations according to the formulation above are executed on a tetrahedral
mesh with about $280\,000$ elements. To show convergence for a higher finite element (FE) order,
the steel plates of the dimension $3.2\,\mathrm{mm}\times 50\,\mathrm{mm}$, see Fig. \ref{fig:structure}, are
coarsely discretized by an edge length of $3.2\,\mathrm{mm}$, the short side of the steel plates.
The first order finite element dicretization contains about $300\,000$ DOF, whereas the second order
discretization results in about $1.8$ mio. DOF on the same mesh.

The numerical results using the explicit Euler scheme in comparison to measurements from
\cite{NakataFuji1990}
are depicted in Fig. \ref{fig:plot_mag_flux}. The quantity under test is the average magnetic flux density
evaluated on the cross sections of the steel plates S1-S3, see Fig. \ref{fig:structure}. Both, for
first and second order finite element solutions a good agreement with the measurements is
noticeable when the magnetic flux saturates. In the time range $0\leq t\leq 60\,\mathrm{ms}$ a more
significant deviation between the lowest order finite element simulation and the measurements is visible. In order to analyse this
behavior a second order FEM simulation using the commercial \emph{CST Studio Suite 2019}
software \cite{CST} is executed. This involves a different tetrahedral mesh of about $2\,000\,000$ elements
where the short side of the steel plates have a finer discretization with four layers of elements. In Fig. 
\ref{fig:plot_mag_flux} the second order finite element solution with the explicit scheme
coincides well with the solution obtained from the \emph{CST Studio Suite 2019} software,
thus it is taken as reference solution for further investigations.

\begin{figure}[]
\centerline{\includegraphics[width=\columnwidth]{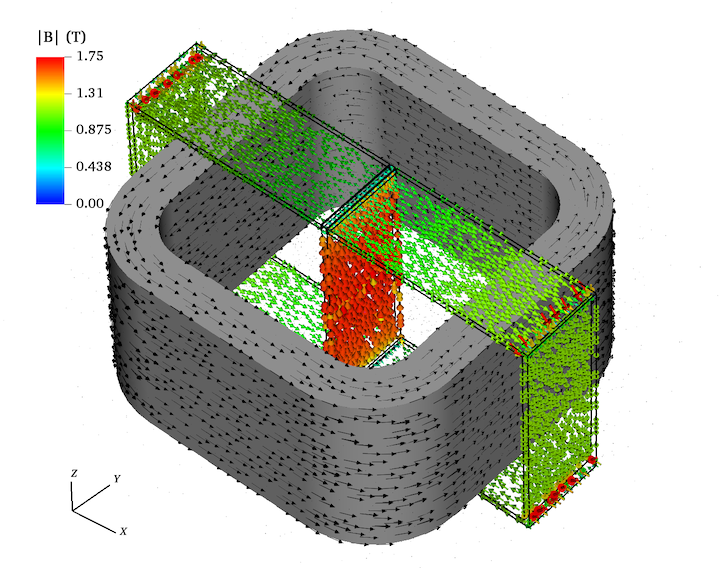}}
\caption{Vector field plot of the magnetic flux density $\vec B$ inside the
steel plates (black wire frame) and homogeneous source current density $\vec J$
(black arrows) inside the coil (gray solid). Second order FEM solution at time $t=150\;\mathrm{ms}$.}
\label{pic:vectorplot}
\end{figure}

The deviations of the first order finite element solution to the reference solution
is explained by a not  sufficient spatial resolution
of the skin effect.
%
In this case study the strongest skin effect occurs at time $t=30\,\mathrm{ms}$, see
Fig. \ref{fig:plt_skin_effect}. The relative error of the explicit Euler solution using 
second order finite elements is less than $5\%$, see TABLE \ref{tab1}. Hence, the 
given mesh resolves the skin effect sufficiently with respect to the error by using
second order finite element discretization.
However, when the magnetic flux saturates less eddy currents are induced. Hence, the magnetic flux is
approximately homogeneous and the coarse spatial discretization by using first order finite elements
less affects the numerical error, which is $1.3\%$ at time $t=150\,\mathrm{ms}$.
Fig. \ref{pic:vectorplot} shows the homogenous magnetic flux density in saturation as second order
finite element solution, where the relative error is less than $1\%$.

\subsection{Performance study of the Runge-Kutta-Chebyshev method versus the explicit Euler scheme}
A sufficient spatial discretization is crucial to resolve effects caused by induction as are the 
eddy currents and the skin effect. Due to the stability criterion (\ref{eq:exlpEulerStab})
of the explicit Euler scheme, the time step size decreases quadratically by refining the
spatial discretization. The RKC method provides a possibility to compensate the decrease of the maximum
stable time increment by choosing a corresponding number of stages $s$, see equation (\ref{eq:RKCStab}).

\begin{figure}[]
\centerline{\includegraphics[width=\columnwidth]{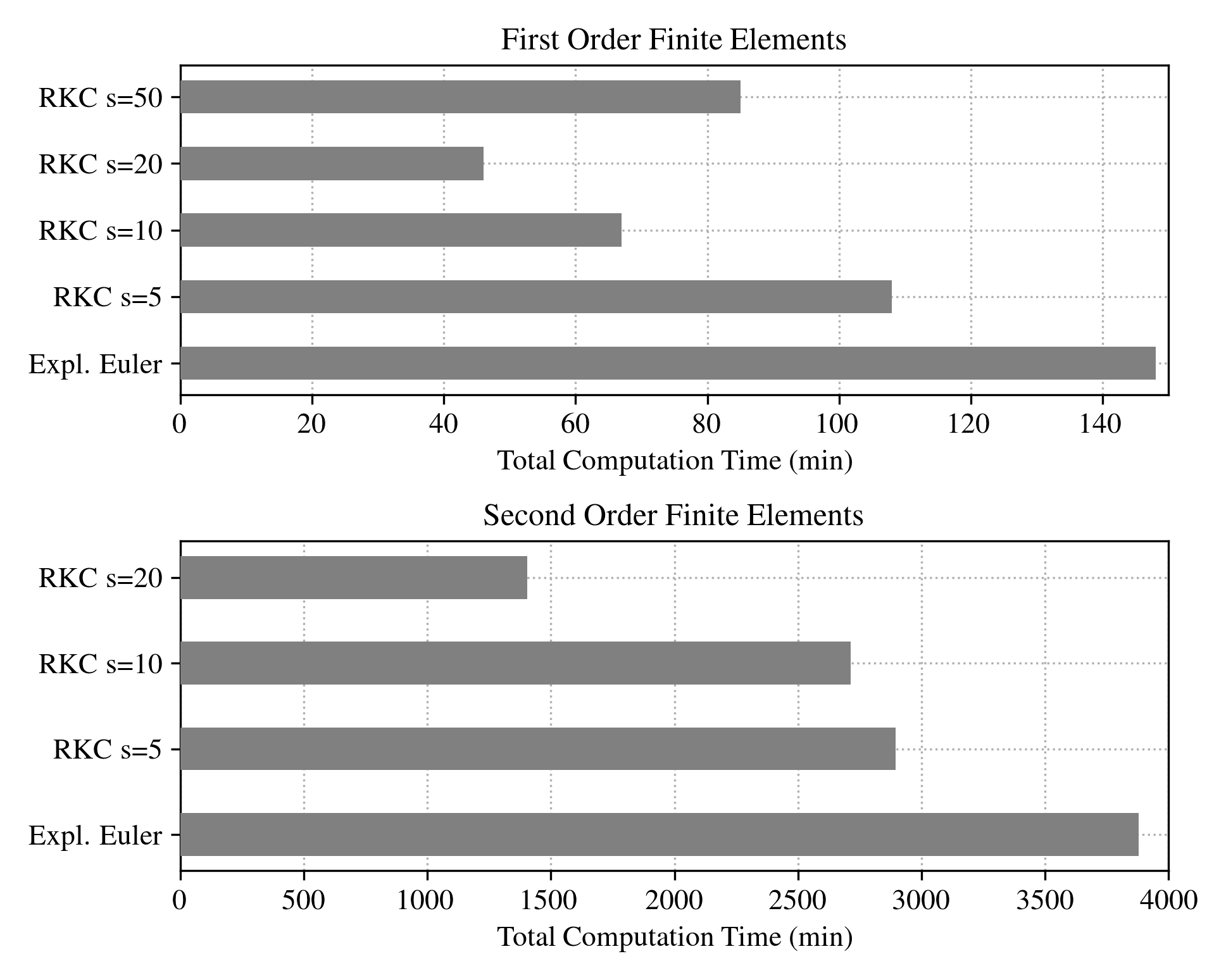}}
\caption{Total computation time of the explicit Euler scheme and the RKC method with different numbers
of stages $s$ for first (top) and second (bottom) finite element order.}
\label{pic:walltime}
\end{figure}

Within the Runge-Kutta stages (\ref{eq:RKC}), where the canonical RHS $\boldsymbol f$ is evaluated
at different time points $t^i+c_j\tau$, $2\leq j\leq s$ and $c_s=1$, the time dependent curl-curl matrix 
$\boldsymbol K_{\mathrm c}(\boldsymbol a_{\mathrm c})$ is kept constant without regarding the update
strategy described in section \ref{sec:update}. The criterion (\ref{eq:update}) is only evaluated after
a full time step.

Fig. \ref{pic:walltime} shows the total computation time of the RKC method in comparison with the
explicit Euler scheme for first and second order finite elements. As expected, in both cases the 
explicit Euler scheme features the longest computation times. Employing the RKC method, the
total simulation time
consumption is diminished significantly for a higher number of stages $s$. However, $s$ has to be
chosen with care as larger time step sizes can trigger the update criterion (\ref{eq:update}) after
each time step, which compromises the overall performance. For this reason, the RKC method with
$s=50$ requires longer computation times than the RKC method with $s=10$ in the first order finite element setting.

\begin{table}[ht]
\caption{Relative error of explicit Euler and RKC simulations to the
implicit reference solution}\label{tab1}
\begin{center}
\begin{tabular}{|c |c|c|}
\hline
& \multicolumn{2}{|c|}{}\\[-2mm]
& \multicolumn{2}{|c|}{$err = \frac{|| B_{\mathrm{ref}} -  B ||_\infty}{|| B_{\mathrm{ref}} ||_{\infty}}$}\\[2mm]
Time Integration Scheme	&			First Order FE&		Second Order FE\\\hline
Expl. Euler&		$0.13352$&			$0.04168$\\
RKC $s=5$&		$0.13347$&			$0.04169$\\
RKC $s=10$&		$0.13349$&			$0.04170$\\
RKC $s=20$&		$0.13354$&			$0.04172$\\\hline
\end{tabular}
\end{center}
\end{table}

The error of all explicit methods with respect to the reference solution is
approximately the same and is mainly given by the error in space.
The deviation in the magnetic flux density is about $13\%$ and $4\%$ for
first and second order finite elements, respectively, see TABLE \ref{tab1}.

All computations are executed with 20 parallel MPI processes on a workstation
with an \emph{Intel Xeon E5} processor. The C++ library \emph{MFEM} \cite{mfem} is involved
for obtaining the finite element operators, the library \emph{PETSc} \cite{petsc} 
provides the parallel PCG solver and the POD method  as well as parallel direct solver.

\section{Conclusion}
For transient nonlinear magneto-quasistatic field simulations, in this paper
a magnetic vector potential formulation was discretized using
the finite element method of first and second
polynomial order and the resulting system of DAEs was transformed into an ODE system
employing a generalized Schur complement. This system of ODEs was integrated
in time by the explicit Euler scheme and an explicit multistage Runge-Kutta-Chebyshev 
formulation of higher order, respectively.

Resolving eddy currents and the skin effect requires a suitably chosen fine spatial
discretization. Since the maximum stable time step size of the
explicit Euler scheme decreased quadratically for linear mesh refinements,
the computation time was increased significantly. To compensate this effect of a
diminished time step size, the RKC method was employed, 
where the maximum stable time step increase quadratically with the number of stages.
The RKC method revealed a significant gain of performance over the
explicit Euler scheme. The numerical tests showed that the performance of
the RKC method is yet limited by the update strategy of the magnetic field
dependent curl-curl matrix.


\section{Acknowledgement}
This work was supported by the Deutsche Forschungsgemeinschaft (DFG)
under grant CL143/11-2 and SCHO1562/1-2. The work of S. Schöps was
supported by the Excellence Initiative of the German Federal and State
Governments and the Graduate School of Computational Engineering, 
Technical University of Darmstadt.


\end{document}